\documentclass{article}
\usepackage{emulateapj}

\def\kms{km sec$^{-1}$}
\def\11{NH$_3$(1,1)}
\def\22{NH$_3$(2,2)}
\def\33{NH$_3$(3,3)}
\begin{document}

\slugcomment{To appear in The Astrophysical Journal}

\title{\center The Dynamics of Molecular Material Within 15 Pcs of the Galactic Center} 
\smallskip
\smallskip
\author{Alison L. Coil}
\affil{Astronomy Department\\
University of California at Berkeley \\
Berkeley, CA 94720 \\
e-mail: acoil@astro.berkeley.edu \\}
\author{Paul T.P. Ho}
\affil{Harvard-Smithsonian Center for Astrophysics \\
60 Garden St. MS-78\\
Cambridge, MA 02138\\
e-mail: pho@cfa.harvard.edu \\}

\smallskip
\smallskip
\begin{abstract}

\ \ \ \ \ We report the results of a 5-field mosaic of the central 15pc
of the Galaxy in the (1,1) and (2,2) lines of NH$_3$.  Two narrow
filaments or streamers are seen running parallel to the Galactic
plane.  The southern streamer appears to carry
gas directly toward the nuclear region from the 20 \kms cloud.  
The eastern streamer, which
we will denote the molecular ridge, appears to be the denser part of
the 50 \kms cloud which lies immediately east of the Sgr A East
complex and extends in the south towards the 20 \kms cloud.  This
ridge of gas carries the kinematical signatures of interactions
with Sgr A East as well as a SNR which lies south of the Galactic
center.  The bulk motion of the gas, the enhanced line widths, and the
heating of the molecular material all suggest an active evolutionary
phase for the gas immediately adjacent to the nucleus.

\end{abstract}


\section{Introduction} 

\ \ \ \ \ The two giant molecular clouds (GMCs) found within 
a few arcminutes of Sgr A$^*$ are known to be physically located at
 the Galactic Center and are seen interacting with the nuclear 
region. As discussed in Coil \& Ho (1999, Paper 1), the 20 \kms cloud 
(``M-0.13-0.08''), lying 10 pc directly south of Sgr A* in projection (using 
R$_{\mathord\odot}$=8.5 pc), appears to be feeding the circumnuclear disk 
(CND) via a molecular gas streamer called the ``southern streamer'' in 
Paper 1. The velocity gradient seen along the southern streamer
places the 20 \kms cloud slightly in front of the CND along
our line of sight.   The 20 \kms cloud is connected to the other
nearby GMC, the 50 \kms cloud (``M-0.02-0.07''), by a ridge of gas and dust
 (Ho et al. 1991; Dent et al. 1993).  This ridge appears to be compressed
gas which wraps around Sgr A East and continues to the core  
of the 50 \kms cloud, located $\sim$ 2' (5 pc) east of Sgr A*.
Sgr A East is an expanding shell of synchrotron emission which lies 
{\it behind} Sgr A* and the CND (\cite{ped89}; \cite{yus86}). It is roughly 10 pc across and appears to be the result of
an explosion which had an estimated total energy of 
$\sim$5 10$^{42}$ ergs (\cite{mez89}; \cite{gen90}), 
over an order of magnitude greater than the energy
released by a typical SNR.  It is unclear at present whether
Sgr A East resulted from a highly energetic SNR, a typical 
SNR which expanded into a pre-existing bubble, or from a nuclear event 
peculiar to its location in the Galactic Center.  It has been suggested
that the tidal disruption of a star by the massive black hole Sgr A* could
account for the energy and morphology of Sgr A East (Khokhlov \& Melia 1996).
 \newline

\ \ \ \ \ Sgr A East is expanding into the 50 km sec$^{-1}$ cloud 
(\cite{gen90}; \cite{hoe91}; Serabyn, Lacy \& Achtermann 1992),
sweeping up gas in the western edge of the cloud, pushing it away 
from the nuclear region and creating a sharp, narrow ridge of dense gas.
Kinematics of the ridge indicate that the gas in the 50 \kms cloud is 
being pushed both to the east and behind Sgr A East along the line of sight, 
which would place the 50 \kms cloud adjacent to and partially behind Sgr A 
East. The velocity shift resulting from this expansion is reported as 
20 \kms by Ho et al. (1991), as seen in north-south position-velocity 
diagrams taken along the western edge of the molecular ridge with a 
spectral resolution of 9.8 \kms.   
Genzel et al. (1990) detect a shift of roughly 40 \kms (resolution $\sim$2.5 
\kms) in an east-west position-velocity diagram taken +40'' in declination 
relative to Sgr A*, and Serabyn 
et al. (1992) (resolution 0.6 \kms) also measure a shift of 40 \kms based
on spectra located just west of the ridge.  
 The line width in the core of the
50 \kms cloud is FWHM=40 \kms, which is unusually broad for molecular cloud 
cores.  The ridge of shocked gas
is long and narrow, roughly 3 pc across and extending well over 10 pc north 
to south.  Along the length of the entire ridge connecting the 20 \kms cloud
with the 50 \kms cloud and along the compression wave front there is a  
roughly north-south velocity gradient of 5-6 \kms arcmin$^{-1}$.  \newline

\ \ \ \ \ Four HII regions are seen along the eastern edge of 
Sgr A East, in the vicinity of the dense ridge of gas at the edge of the
50 \kms cloud.  The HII regions all have velocities of 43-51 \kms 
(\cite{gos85}; \cite{ser92}), and many have shell-like morphologies.  
The HII regions seem to be located in the 50 \kms cloud, close to the
near edge along the line of sight.  An H$_2$0 maser lies 
near the HII regions, possibly resulting from the expansion of Sgr A 
East into the surrounding molecular gas (\cite{yus95}).   Several
1720 MHz OH masers are detected around the Sgr A East shell, both
along the southern rim and on the northwestern side near the CND 
(\cite{yus96}; \cite{yus99}).  The 1720 MHz OH maser line has been
found to be a good diagnostic of shock excitation, being found consistently
at shock boundaries of supernova remnants (SNRs) (\cite{fra96}).
Yusef-Zadeh et al. (1996) propose that the group of masers 
on the southern rim traced the expansion of Sgr A East into the 50 \kms
cloud.  \newline

\ \ \ \ \ Figure \ref{cartoon1} schematically outlines the relevant
large-scale features in the central 15 pc of the Galaxy. Paper 1 discussed
the southern streamer and its interactions with the CND.  
In this paper we report on NH$_3$(1,1) and (2,2) observations
of the 50 \kms cloud and molecular ridge and discuss a SNR to the south
of Sgr A East.  Section 2
details the observations and data reduction, while Section 3 presents
our results, which are discussed in Section 4.  We present our 
conclusions in Section 5. \newline

\section{Observations and Data Reduction}

\ \ \ \ \ We observed the metastable (J,K)=(1,1) and 
(2,2) transitions of NH$_3$ at the frequencies 23.694495 and 23.722633 GHz
using the VLA, operated by NRAO, in DnC configuration.\footnote{ The National 
Radio Astronomy Observatory is a facility of the National Science 
Foundation operated under cooperative agreement by Associated
 Universities, Inc.} A total of five overlapping 2' fields were 
mapped towards Sgr A* on the night of 1995 February 10. 
The mosaiced field of view is $\sim$ 4' by 5', covering the inner 10-12 pc 
of the Galaxy. Results from the western three fields are presented
in an earlier paper (Paper 1), and here we will show velocity-integrated
images from all five fields but will focus mainly on the eastern
two fields, which cover the western edge of the 50 \kms cloud and the 
molecular ridge.  Our
total bandwidth is 12 MHz centered at $v_{LSR}$= 31.14 km $s^{-1}$ for the
eastern two fields. The spectral resolution is 
4.9 \kms, corresponding to a total velocity coverage of 140 \kms for our
final 28 channels.  
We tapered our images to a synthesized
beam size of $\sim$ $14''\times9''$ with a position angle of 
12.6$^{\circ}$.  
Details of the flux, phase and bandpass calibrations are 
described in Paper 1.   
Data from all five fields were jointly imaged and deconvolved in MIRIAD,
effectively mosaicing the data in the u,v plane.  We employed the same
CLEANing technique as in Paper 1, where we added a flat offset of 
zero-spacing flux during the CLEAN process which was removed after CLEANing.
 The images presented in this paper have all been corrected for the 
primary beam response unlike many of the
images shown in Paper 1 which were tapered at the edges, resulting in a lower
correction (and lower noise) at the edges of those images.  All coordinates
are B1950 epoch. \newline

\section{Results}

\subsection{Velocity-Integrated Emission}

\ \ \ \ \ Figure \ref{color.mom0} is an NH$_3$ velocity-integrated 
mosaic map of all five fields seen in contours overlaid on a 
color 20 cm continuum image of Sgr A East and West (Lang, Morris, \& 
Echevarria 1999).  Sgr A West is the over-saturated black region, 
inside of which the mini-spiral of ionized gas streamers converge near the 
location of Sgr A*.  The red emission traces Sgr A East, which is offset 
from Sgr A West. 
The contours display \11 emission on the left and \22 on the right. 
There are two prominent north-south streamers in these images.  
The ``southern streamer'' (see Paper 1) 
connects the circumnuclear disk (CND) surrounding the 
ionized gas mini-spiral with the 20 \kms to the south.  The northern 
half of the 20 \kms cloud is shown in these images, located at
R.A.=17$^h$42$^m$28$^s$.5, Dec.=-29$^\circ$02$'$.0. 
As reported in Paper 1, we find a velocity gradient 
of $\sim$5-8 \kms arcmin$^{-1}$ along the southern streamer with the gas 
redshifting to the north, which places the 20 \kms cloud in {\it front} 
of the nuclear region along the line of sight.  Details of this streamer 
are discussed further in Paper 1. \newline

\ \ \ \ \  The other long streamer in Figure \ref{color.mom0} is the 
molecular ridge which traces the denser parts of the 50 \kms cloud, including
 the core of the cloud and gas in the southern part of the GMC 
 where Sgr A East is expanding into the cloud.  
There is strong emission in the northern half of the streamer where the
ridge wraps around the eastern edge of Sgr A East.  The 50 \kms cloud 
continues to the east and north beyond the edge of the figure 
(Zylka, Mezger, \& Wink 1990; \cite{ser92}; \cite{den93}).
The narrow, concentrated ridge of molecular gas wraps around Sgr A East
and continues to the south at a lower flux level, past -29$^\circ$03$'$ to the 
edge of our sampled field.  
The width of the molecular ridge is roughly constant along its
length, though to the south of Sgr A East the emission is fluffier and more
spread out. The gas is highly clumped, 
and the sharp edges where the intensity of the emission increases 
quickly are not smooth along the length of the ridge. The emission in the
northern part of the ridge falls off more steeply at the edges
than in the southern part, which may indicate that the gas to the north
is more confined than to the south.  The northern
part of the ridge closely follows the arc of Sgr A East, appearing to wrap
around the expanding shell.  There are many small bumps and protrusions 
of the gas which are seen in both the \11 and \22 data, 
including the location of the peak emission in the northern part of
the ridge. The flux in the ridge is higher in the north, and
the peak is found lying along the easternmost part of Sgr A East. \newline

\ \ \ \ \ There is an extension of the gas projected onto 
Sgr A East near Dec.=-28$^\circ$59$'$0.5$''$, where there is a 
large hollow bubble in the continuum emission greater than 1$'$ in 
size, clearing gas into the molecular ridge to the east and Sgr A West 
and the CND to the west.  This hollow bubble can be seen clearly in Figure
\ref{color.mom0} where there is a yellow hollow region inside the red emission
from Sgr A East, near the southeastern edge of Sgr A East in between the 
two streamers.  The molecular emission closely traces
the eastern and western edges of the bubble, wrapping just around 
the outside of the hollow area.  Morphologically it appears that
the bubble is interacting with both the molecular ridge and the gas in
the northern part of the streamer connecting the 20 \kms cloud to the 
CND (the ``southern streamer'' in Paper 1).
To the south of this bubble, there is another region where the gas from
the ridge has a protrusion to the west, near the southern edge of Sgr A 
East.  At this location a connection has been imaged between 
 the molecular ridge and the 
20 \kms cloud in NH$_3$(3,3) (\cite{hoe91}).  We do not image a 
connection here, but connections between these features have
also been imaged in   
$^{13}$CO(2-1) (\cite{zyl90}) and submillimeter thermal dust
emission (\cite{den93} and \cite{zyl97}).  \newline

\ \ \ \ \ Along the eastern edge of Sgr A East, the molecular ridge 
continues without break or bend to the south past where the continuum 
emission stops.  The separation between the
two streamers in the southern half of these images is roughly constant,
which may in part be affected by the deep negative sidelobes due to the
lack of short spacings and the fact that the mapped region is sampled only
at full-beam spacings.  
However, the narrowness of the ridge in the southeast is not likely due
to the primary beam response as our 
primary beam includes much of the empty region in the southeast as seen in 
Figure \ref{beam.mom0}.  This image shows our effective mosaiced
primary beam pattern on the sky and labels the percentage level of 
flux sampling over the region. We have not plotted emission with a
gain below 15\% as those features are unreliable.
While the beam response is not uniform across the field, it is at least 50\% 
everywhere, so that at most our flux may be down by a factor of two in
the very central region. 
The northernmost part of the molecular ridge shows high flux near the
edge of the primary beam, indicating that the ridge continues to the
north beyond our sampled field.  Our limited velocity range may also
cause undersampling in this region.  An area which is interesting
for its lack of flux is the lower left of the map, to the east of the
southern part of the molecular ridge, where with a 100\% beam response 
level we do not detect emission. \newline

\ \ \ \ \ Many of the most striking features seen in Figure \ref{color.mom0}
are apparent in both \11 and \22 emission. Similarities between the images
include the narrow width of the molecular ridge, the location of the 
 peak emission in the ridge, and the extension of gas 
into Sgr A East at the eastern side of the bubble. The localized emission 
in between the two streamers at the southern edge of Sgr A
East (Dec.=-29$^\circ$00$'$30$''$) is seen in both maps, and the 
structure of the 20 \kms cloud is very similar in the \11 and \22 
images, where the many peaks and fingers of emission are the same.
The main differences between the \11 and \22
velocity-integrated emission maps are seen concentrated near 
the nuclear region (discussed at length in Paper 1). \newline

\ \ \ \ \ Figure \ref{bw.SNR.mom0} shows the same continuum image
as Figure \ref{color.mom0} with our velocity-integrated NH$_3$ 
contours, but here the stretch of the grey-scale has
been altered to bring out lower level features in the 20 cm continuum
image.   Four small HII regions are located just 
off the eastern edge of Sgr A East, seen here as small black spots 
at the top left of the continuum image (\cite{eke83}; \cite{gos85}).  
The molecular ridge of gas is narrowly confined to the region 
in between the eastern edge of Sgr A East and the HII regions 0.5$'$ away.
In the \22 image the northernmost part of the streamer above 
Dec.=-28$^\circ58'$30$''$ 
is narrower and is
seen just at the edge of Sgr A East and does not extend eastward 
to the northern two HII regions.
The southernmost HII region, located at R.A.=17$^h$42$^m$41$^s$,
Dec.=-28$^\circ59'$12$''$ and labeled D in discussions
of this HII region grouping (\cite{gos85}), 
appears to be still embedded in dense molecular 
gas where the contours peak sharply into a small clump on the eastern edge
 of the ridge.  This HII region is the only one of the four which does not
have a shell-like morphology, which could indicate that it is in an earlier
stage of development. It may either be younger than the other HII regions 
or its expansion or flow out of the cloud may have been slowed by the 
dense gas surrounding it.  \newline

\ \ \ \ \ In Figure \ref{bw.SNR.mom0} there is an 
intriguing association between the continuum and molecular 
line emission along the 20 \kms cloud in the southwestern part of the
image.  The 20 \kms cloud follows the arc of the two dark streaks 
of continuum emission known as Sgr A-F and Sgr A-E (the `wisp') 
(\cite{ho85}; \cite{yus87}).  Ho et al. (1985) suggest that
these two non-thermal arcs trace the edge of a SNR which may be expanding
into the 20 \kms cloud, compressing the gas in the GMC.  Figure 
\ref{bw.SNR.mom0} reveals a hollow shell extending across the
southern half of the image, including these arcs to the southwest
and continuing up around the southern edge of Sgr A East and extending
out along another curved arc centered at  R.A.=17$^h$42$^m$40$^s$,
Dec.=-29$^\circ$00$'$30$''$.  The southern
edge of Sgr A East has a curved shape seen in Figures \ref{color.mom0} 
and \ref{bw.SNR.mom0} which follows the arc of the SNR, 
opposing the curvature of the roughly spherical Sgr A East shell itself.
\newline

\ \ \ \ \ In the \33 velocity-integrated map (\cite{hoe91}, see Figure
1) the molecular ridge does
not continue as far to the south as it does in the \11 and \22 maps.
The \33 image shows the ridge following the eastern arc of Sgr A East
down to Dec.=-29$^{\circ}$00$'$30$''$ where the ridge then splits off
from Sgr A East and continues to the southwest towards the 20 \kms
cloud.  The ridge does not continue south of Dec.=-29$^{\circ}$02$'$.
In the \11 and \22 images, the ridge has the same behavior as the
\33 emission down to Dec.=-29$^{\circ}$02$'$,
where instead of connecting with the
20 \kms cloud as in the \33 image, it continues further south to
Dec.=-29$^{\circ}$03$'$.  Because the length of the southern part of 
the ridge is different from the \33 image, we investigate this
emission further to look for possible contamination by sidelobes from 
the 20 \kms cloud.  Figure \ref{sgra12.mom0} displays
velocity-integrated maps of the southern
two fields of the \11 data imaged individually without any mosaicing
of other fields.  The image on the left is from the southeastern
field only and does not show any emission from the 20 \kms cloud.
The southern part of the ridge can not be due to sidelobe structure from
the 20 \kms cloud as the ridge is seen clearly in the primary beam 
centered on this region. 
The southwestern field is shown on the right, where the 20 \kms cloud
and part of the southern streamer is seen.  The negative
contours are due to sidelobe structure, as are the positive contours
in the eastern half of the image.  This figure shows that the southern
half of the ridge is not strongly affected by sidelobe structure
from the 20 \kms cloud.  We also show position-velocity cuts taken
along the southern part of the ridge and the 20 \kms cloud in Figure
\ref{sidelobe} to see if the kinematics of the gas in the ridge and 
the GMC are identical, as would result from strong sidelobe contamination.
The position-velocity diagrams from cut A do not match those from cut
B in either \11 or \22 emission.  The locations of the 
velocity peaks, as well as the line widths, do not correlate between
the two cuts.  We conclude that the structure of the southern part
of the ridge is real and that contamination by sidelobes
from the 20 \kms cloud is not important. \newline

\subsection{Comparison with 1.3mm Dust Emission}

\ \ \ \ \ Figure \ref{dust.mom0} overlays contours of 1.3mm emission
(\cite{zyl97}) in the Sgr A complex with a grey-scale image of 
our \11 velocity-integrated emission.  The 1.3mm map traces both
free-free emission from ionized gas in the Sgr A West HII region
and thermal dust emission from the GMCs which traces the total column of gas.
The 1.3mm data has an angular resolution of 11$''$ and is quite comparable
to the resolution of the NH$_3$ data.  As single dish data however,  the
1.3mm image is more sensitive to emission on larger scales than our
interferometric data, which are missing the shortest spacing information,
 so that the 1.3mm map better traces the large-scale structure of
the material in the region.  The peak of the 1.3mm emission delineates 
the ionized gas in the  Sgr A West HII region 
where the southern half of the inner edge of the CND can be seen in 
grey-scale at the lower edge of the spiral.  The streamer
to the south which connects the nuclear region with the 20 \kms cloud is
present in both the dust and molecular gas images, where the 1.3mm contours 
closely trace the \11 emission.   The core of the 20 \kms cloud is seen
at the lower edge of the 1.3mm image, beyond the spatial 
coverage of our \11 map.  The dark grey-scale emission traces only
the northern part of the 20 \kms cloud, where the GMC is becoming
narrower and elongated, connecting with the nuclear region.  There is
extremely excellent agreement in detailed spatial structures between the
\11, \22, and dust maps.  The other
dark region of the grey-scale \11 map marks the core of the 50 \kms 
cloud in the upper left of the image, near RA=17$^h$42$^m$40$^s$
and Dec.=-28$^{\circ}$59$'$.  The entire 50 \kms GMC includes 
fluffier emission not traced by the NH$_3$ emission in
the molecular ridge, and extends to the north beyond the extent of
our images. Near the 50 \kms cloud core and towards the south where 
the 50 \kms cloud bends towards the Sgr A West spiral,  the \11 
emission in the molecular ridge closely corresponds to the areas of
 highest dust column
density as seen in the 1.3mm image.  However, at the cloud core 
the \11 emission does not continue to the eastern 
edge of our field where the dust column density is still high.  \newline

\ \ \ \ \ It has been suggested that the 50 \kms cloud may also feed
the CND via a streamer of gas and dust in a similar manner as the 20 
\kms cloud. An intriguing gas streamer has been seen in single-dish 
HCN emission by Ho (1994), where a distinct narrow streamer appears
to connect the western edge of the 50 \kms cloud with the nuclear region.
This streamer is seen well in the individual channel maps, specifically 
around 60 \kms.  A similar streamer connecting the 50 \kms cloud
with the CND has been imaged in $^{13}$CO(2-1) emission by Zylka et al. (1990).
The streamer can be seen in the channel map showing emission from 
50 \kms to 60 \kms (their Figure 5.d).  This streamer has not been
seen in other millimeter and sub-millimeter observations (\cite{mez89}; 
\cite{den93}) nor in interferometric NH$_3$(3,3) data (\cite{hoe91}), 
nor do we detect it here.  It remains unclear whether the 50 \kms cloud
 directly feeds the nuclear region or whether the streamer is being 
projected along the line of sight, though it is clear from the dust
maps that if an eastern streamer exists it must have a lower column
density than the southern streamer.  There is also an intriguing 
finger of dust emission to the north of the 50 \kms
cloud which points in towards
the nucleus, at the top of the CND near RA=17$^h$42$^m$33$^s$ and
Dec.=-29$^{\circ}$58$'$. \newline

\ \ \ \ \  The \11 molecular ridge continues to 
the south of the 50 \kms cloud core where the dust emission is  
lower (Dec.=-29$^{\circ}$00$'$ to -29$^{\circ}$03$'$).    
The ratio of flux in the 50 \kms cloud and the  20 \kms cloud to the 
southern part of the ridge is higher in \11 than in 1.3mm dust emission. 
 This implies that the relative prominence of the \11 emission in the 
southern part of the ridge as compared to the 1.3mm emission is due in 
part to the contouring of the 1.3mm image and is most likely due to
 the sensitivity of the 1.3mm map to extended lower density material.
The long molecular ridge as seen in the \11 image has a constant,
narrow width which is reflected in the dust emission only along the
northern half of the streamer. 
At the southern tip of the molecular ridge there is a 
U-shaped feature in the 1.3mm image connecting the ridge with the
eastern side of the 20 \kms cloud.  A connection between the 
two GMCs has been imaged in NH$_3$(3,3)
(Ho et al. 1991), with similar coverage as our \11 and \22 maps, and in 
submillimeter continuum (Dent et al. 1993), which has similar coverage as 
the 1.3mm emission shown here.  It appears from Figure \ref{dust.mom0}
that the two GMCs are not entirely separate entities connected only
by a thin ridge of gas, but the clouds and ridge may be intimately 
connected and part of a coherent larger structure 
of gas lying along the Galactic plane.  \newline

\ \ \ \ \ Striking features of our velocity-integrated \11 and \22 images
include the narrowness of the two streamers seen lying along the Galactic 
plane and the roughly constant separation between them.
Our lack of zero-spacing information prevents us
from concluding whether the narrowness of the streamers is meaningful.
The roughly constant spacing between the two streamers in our maps
 is not seen in other images of the region and may be due to 
the negative trenches resulting from sidelobe structure in between the
two streamers (see Figure \ref{color.mom0}).
Comparing our maps to the \33 and 1.3mm
emission it seems that we are missing structure in between
the two streamers at the southern edge of Sgr A East.
The total column density as seen by the dust emission shows the molecular
ridge curving around Sgr A East in the south, pointing towards the 
southern streamer as it approaches the nuclear region.  
The \33 map more closely follows the dust emission in this region than 
our \11 and \22 images, and in the \33 image there are
two connections seen between the streamers: one along the southern
edge of Sgr A East, which is also seen in the dust emission,  and another
connection further south at Dec.=-29$^{\circ}$02$'$.  Our images do not 
show either of these connections, which may be a result of the negative
sidelobe trench or may also be a temperature effect. 
The 1.3mm dust emission does show low-level flux along the
southern part of the molecular ridge  but it 
is not confined to a narrow streamer, whereas the northern half of the 
ridge is narrow in both our maps and the 1.3mm map. \newline

\subsection{\22/\11 Line Ratio Map}

\ \ \ \ \ Figure \ref{22.11.mom0} indicates regions of heating
as derived from the \22/\11 line ratio where the dark regions
trace heated gas.  
The rotation temperature of the gas can be derived from the ratio 
of \22 to \11 observed brightness temperatures if the lines are 
optically thin.  In the optically thin limit a \22/\11 ratio of
0.5 implies a rotation temperature of 21 K, a ratio of 1 corresponds
to 33 K, a ratio of 1.5 corresponds to 49 K, and a ratio of 2 corresponds
to a rotation temperature of 73 K.  These values are in actuality
a lower limit on the rotation temperature, because if line opacities are 
substantial a correction must be made based on the optical
depth of the \11 line (see Figure 4 of \cite{hoe83} for details of how 
the rotation temperature depends on the optical depth).  
Nevertheless, a simple map of the \22/\11 line
ratio can still provide heating information if the optical depth of the
gas does not change significantly over the region.  
The contours in Figure \ref{22.11.mom0} are from the 6 cm continuum map of 
Sgr A East as seen in Figure \ref{color.mom0}, overlaid on a grey-scale map of 
the velocity-integrated
\22/\11 ratio, with the dark regions of high \22/\11 ratio
 showing hotter gas. Heating in the nuclear region, near the CND,
was discussed in Paper 1.  Here we focus on the molecular ridge
and the 20 \kms cloud, where most of the heating is along the 
western edges of these features.   
There is not much heating seen along the northern part of
the inner edge of the molecular ridge  where the 
gas appears to be directly interacting with Sgr A East.  We are
unable to estimate optical depths at this location from our spectra
and can not determine whether this apparent lack of heating is real.   
 The extension 
of gas projected onto Sgr A East shows signs of heating, as does
the small region of emission seen in between the two streamers
at the southern edge of Sgr A East.  The southern part of the molecular
ridge shows heating on both the eastern and western sides, much like
the 20 \kms cloud emission.  Much of the heating in this map is along
the edges of the streamers.  This is not an imaging artifact, where the
edges are darker due to cutoffs in the processing of the maps, as \22/\11 
ratio maps made with many different cutoffs show all of the same 
regions of heating. The apparent edge heating may well be affected by 
optical depth differences along the edges of the streamers. \newline

\subsection{Velocity Centroid and Dispersion Maps}
 
\ \ \ \ \ Figures \ref{mom1} and \ref{mom2} display
kinematic features of the gas in the molecular ridge as well as the southern
streamer.  Figure \ref{mom1}
maps the mean velocity of the gas at each pixel, where the dark regions
correspond to higher velocity gas. The grey-scale ranges from 15 \kms for
the white areas to 60 \kms in the black regions.
 There is an overall velocity gradient
along the length of the molecular ridge, where the gas in the northern 
half of the ridge has been redshifted by $\sim$20 \kms.
The mean velocity does not appear to increase linearly along the length of
the streamer, but instead the northern half of the streamer appears to be
at $\sim$40 \kms where the ridge wraps around the edge of Sgr A East,
while the southern part of the streamer also appears to be
at $\sim$20 \kms.  This change in velocity near the southern edge of Sgr A
East, where the ridge does not appear to be directly interacting with 
Sgr A East, can be seen in both the \11
and \22 maps. There are also noticeable smaller dark regions of pronounced 
redshifting of the gas.  This occurs at the inner edge of the 
molecular ridge where the gas appears to extend into Sgr A East
 in projection. This gas is also heated, as seen in 
Figure \ref{22.11.mom0}, and may be located at the far side of Sgr 
A East, being redshifted along the line of sight by the expansion
of Sgr A East. The redshifting of the extension is seen to a lesser 
degree 
in the \22 image, where the gas is highly redshifted at the base of 
the extension but the mean velocity decreases further along the 
extension.  There is another intriguing localized region of 
redshifting seen in the \11 image at the
outer edge of the molecular cloud, near Dec.=-29$^{\circ}$00$'$45$''$.  
The eastern edge of the 20 \kms cloud is also redshifted relative
to the rest of the cloud, and at the lower tip of the cloud there is 
a small area of redshifted gas where there are a few contours from
the 6 cm continuum map.  This non-thermal feature, dubbed the 
``wisp'' (\cite{ho85}), has been proposed to be the result of a 
supernova remnant
interacting with the ambient molecular gas.  The redshifting
of the molecular gas in the 20 \kms cloud at this location as
seen in Figure \ref{mom1} may be explained by this idea. \newline

\ \ \ \ \ Figure \ref{mom2} spatially maps the velocity dispersion of the
molecular gas in grey-scale, where darker regions indicate gas with 
higher dispersion.  The darkest regions around Sgr A West are where
the gas from the southern streamer is accreting onto the CND (Paper 1).
The southern tip of the 20 \kms cloud shows high velocity dispersion
at the ``wisp,'' which further strengthens the claim that the ``wisp''
is the result of a SNR (\cite{ho85}).  Along the western molecular ridge,
there is little change in dispersion in the \11 image 
across the northern half of the ridge from east
to west where the ridge wraps around the eastern side of Sgr A East,
 possibly indicating that all of this gas has been processed
and is in a post-shock phase, which is consistent with the lack of heating
here (see Figure \ref{22.11.mom0}).  In the \11 image there are two small 
darker spots of higher
dispersion along the northeastern edge of the ridge
which may be the result of infall or 
outflows from the HII regions seen in contours. 
The \22 image shows some line broadening along the inner edge of the
northern part of the ridge facing Sgr A East, where the gas is darker
along the western side of the ridge.
As in the velocity centroid map (Figure \ref{mom1})
 there is a demarcation between the upper and lower halves
of the molecular ridge, where wide variations in dispersion are
seen throughout the southern half of the ridge.  
In particular, there are concentrated regions of increased velocity 
dispersion along the inner,
western edge of the ridge which are not seen in the northern half.
The gas in the lower part of the ridge generally appears to be more 
perturbed than the gas directly adjacent to Sgr A East.  
Many of the dark regions in the \22 map also show unusual mean velocities 
in Figure \ref{mom1} and are heated as seen in Figure \ref{22.11.mom0}.  
Clearly the gas along the lower inner edge of the molecular ridge is
 highly perturbed.  This is consistent with the SNR to the south of Sgr A 
East impacting the gas.  \newline

\subsection{Position-Velocity Diagrams}

\ \ \ \ \ We further investigate the kinematics of the gas using
position-velocity diagrams in Figures \ref{psvl1} and \ref{psvl2}.  
Figure \ref{psvl1} displays position-velocity information for \11
and \22 emission as well as the \22/\11 ratio 
 along four cuts running the length of the molecular ridge.
The datacubes have been smoothed in R.A. and Dec. to enhance the
general spectral features and increase the signal-to-noise.  The
\22/\11 datacube has also been Hanning smoothed in velocity space.
The diagrams for cut A from the outer side of the streamer, furthest 
from Sgr A East, show a velocity gradient along the southern part of
the streamer starting at 20 \kms at the southern tip of the streamer
and increasing past 55 \kms towards the middle of the streamer.  
The velocities then decrease again to the north at the location on
the velocity-integrated map where the most emission is found.  The overall
gradient along the ridge traces out a backwards ``C'' shape in 
position-velocity space in the \11 diagram.
The \11 diagram shows a large line width and blueshifted gas at the
northern edge of the cut near the southernmost HII region in the 
continuum image (Figure \ref{color.mom0}).  This
blueshifted emission may be the result of outflows from this HII region.
Further south along cut A at the location of the dark emission on the 
eastern edge of the ridge as seen in \11 emission in 
Figure \ref{mom1}, there is an asymmetric lobe of 
redshifted emission extending from 30 \kms to 80 \kms in the \11 
position-velocity diagram and from 40 \kms to 70 \kms in the \22 diagram.  
This gas is not only redshifted, but has a higher velocity dispersion
than the gas towards the south 
as seen in both the \11 and \22 diagrams.  This broader gas may
be similar in nature to gas further north.  (The \11 gas should be slightly
more spread out in velocity space due to an unresolved
 hyperfine line located 8 \kms away from the main line, which causes 
some artificial line broadening.  Our velocity 
resolution is able to resolve the hyperfine structure in the \22 gas 
so that the \22 diagrams do not show artificially broadened line widths.  
However, \11 emission with line widths across 50 \kms can not entirely
be due to hyperfine blending.)  The \22/\11 diagram indicates heating,
where dark regions trace hotter gas.  This redshifted gas shows heating
for the component at 35-40 \kms, but not for the highly redshifted gas.
An increase in line width is also seen 
at the southernmost tip of the streamer, where there is low-level emission
extending from -10 \kms to 50 \kms and from 65 \kms to 85 \kms.  
This gas appears to be highly perturbed, and the \22/\11 diagram
shows that the component of this gas at high velocities is heated.  \newline

\ \ \ \ \ Cuts B and C, taken through the central part of the ridge,
show a similar overall velocity gradient as cut A, though the 
gas does not trace out a backwards ``C'' pattern in position-velocity space 
as it does in cut A.  The global velocity gradient is of order 7 \kms 
arcmin$^{-1}$, in good agreement with the gradient found by Serabyn et al. 
(1992).  In the \11 diagrams there is some low-level high-velocity
emission in the central region, at velocities of 70 \kms to $\geq$100 \kms.
In the \22 diagrams there is 
redshifted gas which continues as a feature at 70-80 \kms along the southern
length of the ridge, extending in cut C to velocities $\geq$100 \kms.  
The \22/\11 diagrams for B and C  show that this highly redshifted 
gas at $\geq$70 \kms is heated.  There is also heating in the low-velocity
($\leq$0 \kms) gas at the southern end of the ridge, where the velocity 
dispersion increases. The \11 position-velocity diagram for cut D, 
which is taken along the inner edge of the ridge, where the streamer 
is adjacent to Sgr A East, looks very similar to the \11 diagram for
cut A.  The backwards ``C'' shape is again apparent in the northern half of
the diagram, where the gas in
the middle of the streamer has been redshifted to a central velocity 
of 60 \kms as in cut A, whereas in cuts B and C the central  velocity is
45 \kms.  In the \22 diagram for cut D this part of the streamer is
also highly redshifted, with a very pronounced kink at 50 arcsec on
the y-axis.  Gas in the southern part of cut D has a very large dispersion
as seen in the other diagrams, covering our entire sampled velocity range
from below -35 \kms to greater than 100 \kms. There appears to be a hole 
in position-velocity space in the southern half of the diagrams from 0 
arcsec to -75 arcsec on the y-axis and from 40 \kms to 70 \kms on the x-axis.
At the location of this hole the central velocity of the gas is at $\sim$30
\kms and has a narrow line width compared to the gas elsewhere along
the streamer. The \22/\11 diagrams indicate that the gas at the edges of 
the hole is heated.  \newline

\ \ \ \ \ Position-velocity diagrams taken east to west across the
molecular ridge are seen in Figure \ref{psvl2}. We again show diagrams
for both \11 and \22 emission, as well as the \22/\11 ratio to show 
the heated gas.  The \11 emission for cuts A and B,
 taken across the northern section of the ridge, show a velocity gradient
 from low velocities along the eastern edge of the ridge to higher
velocities along the western edge, facing Sgr A East.  The gas in
the east is seen at velocities as low as -5 \kms whereas in the west,
$\sim$3 pc away in projection, gas is seen at velocities up to 85 \kms. 
 The \22/\11
diagram for cut A shows that the redshifted gas to the west is heated,
as seen from the dark region at 70 \kms.  The \22 diagram for cut B, which
intersects the region of the streamer which extends into
Sgr A East in projection, shows an ``O'' shape of emission in 
position-velocity space.  In the east the central velocity of the gas
is 40 \kms, whereas along the extension the gas is both red- and 
blueshifted
by $\sim$40 \kms before it combines again at a central velocity of 35 \kms.
Both Genzel et al. (1990) and Serabyn et al. (1992) report a velocity 
shift of $\sim$40 \kms
in the molecular ridge where it is interacting with Sgr A East.  
The \22/\11 diagram shows that the gas in the west at 35 \kms is strongly 
heated.  This gas is being impacted by the expansion of either 
Sgr A East or the hollow bubble seen in Figure \ref{color.mom0} 
into the ridge, causing red- and blueshifting of the
gas.  The redshifted emission has
a higher flux level than the blueshifted emission, indicating
that more of the gas is located {\it behind} Sgr A East than in front 
of it. \newline

\ \ \ \ \  Cut C is taken in the middle of the ridge 
near the southern edge of Sgr A East where
the ridge no longer wraps around Sgr A East but continues to the south.  
Here the gas in the ridge does not have the 
east to west redshifting velocity gradient, nor is the ``O'' shape
apparent.  The \22/\11 diagram shows that the gas is heated on
the eastern and western edges of the ridge, and that in the center
of the ridge the gas at high velocities ($\sim$75 \kms) is also heated.
Cuts D and E, taken from the southern part of the streamer, both
indicate that the gas in the south has a higher velocity dispersion than the
gas to the north.  There is low-level emission 
at all velocities in our sampled velocity range, from $\leq$-35 \kms up
to $\geq$100 \kms.  The high-velocity gas ($\geq$70 \kms) is heated 
as seen in the \22/\11 diagrams.  The very low-velocity gas ($\leq$-10 \kms)
is also heated, though to a lesser degree.
In opposition to the redshifting velocity gradient seen in
 the northern part of the ridge, here in the south the gas blueshifts
 slightly towards the western edge of the streamer facing Sgr A East.
The western edge of the streamer is heated as seen in cut E,
with a central velocity of 20 \kms.  \newline

\subsection{Derived Physical Parameters}

\ \ \ \ \ We estimate that we are recovering 20-25\% of the single dish \11 
flux in our interferometric data by comparing the peak observed brightness
temperature in each streamer with the single-dish antenna temperatures 
(corrected for beam efficiency and atmospheric attenuation) at the 
same locations reported by Armstrong \& Barrett (1985), after convolving
our data to the 1.4$'$ beam of the Haystack data.  We are not sensitive
to structures greater than 1$'$ in extent, nor do we have short spacing
information which would increase our total flux.  We list derived physical 
parameters for the \11 peak flux locations in the northern and southern parts 
of the ridge in Table 1.  A mean optical depth was derived for each part of
the ridge from spectra with clear hyperfine structure.  A total of 8 and 10
 spectra were used to determine the optical depth in the northern and 
southern parts of the ridge, respectively.  To derive the physical parameters
we used the equations outlined in Paper 1, section 3.6.  An estimation of
the rotation temperature requires the \11 optical depth and the ratio of
\22 to \11 brightness temperatures at that location.  Masses for each part of
the ridge are estimated from the derived column density, which depends on
 the optical depth and excitation temperature.  These mass estimates for
the dense component of the ridge are reasonable even though we are not 
imaging all of the flux as the missing flux corresponds to more extended 
structures and may include some of the 50 \kms cloud.
We find a mass for the molecular ridge of 1.5 10$^5$ M$_\odot$, 
which agrees quite well with masses for the molecular ridge 
calculated by Zylka et al. (1990) and Serabyn et al. (1992).  \newline

\section{Discussion}

\subsection{Interaction Between Sgr A East and the 50 \kms Cloud}

\ \ \ \ \ The northern part of the molecular ridge traces the densest 
parts of the 
50 \kms cloud, including the core and the ridge of dense gas being 
impacted by Sgr A East.  The ridge may be processed and post-shock,
as there is not a significant increase in density or temperature
 along the western edge of the ridge (see Figures \ref{color.mom0} 
and \ref{22.11.mom0}) as one 
might expect if the inner edge was being more directly impacted 
by Sgr A East than the outer edge.  There is also not a pronounced increase
in velocity dispersion along the inner edge of the ridge in the \11 emission,
though in the \22 data the velocity dispersion does increase somewhat
along the inner edge (see Figure \ref{mom2}).  If the ridge is processed
and the shock wave has reached the far side of
the ridge where the string of HII regions is, it is possible that the
shock wave itself created the collapse which formed the HII
regions. Yusef-Zadeh  \& Mehringer (1995) find an H$_2$O
 maser associated with one of the four
sources in the HII region which could indicate that the shock from the
expansion of Sgr A East has reached the HII regions. \newline

\ \ \ \ \  There is an increase in the 
central velocity of the gas along the inner edge in the northern part of 
the ridge, as seen in the position-velocity diagrams for cuts A and B in
 Figure \ref{psvl2}.  
The position-velocity diagram for the \22 emission across the densest part
of the ridge, near the southernmost HII region (cut B), shows both red- and
blueshifting of the gas.  More gas is 
redshifted than blueshifted, indicating that as the ridge wraps around
the expanding shell, it is more behind Sgr A East than in 
front of it (evidence that at least part of the gas cloud lies in front 
of Sgr A East is also seen by \cite{lis83}).  These diagrams include emission
from the small gas extension which is projected onto Sgr A East.  In
the \11 diagram, this gas is redshifted to 55 \kms and 75 \kms, while
in the \22 diagram an ``O'' shape is seen.  This ``O'' shape would
most naturally result from an expanding bubble.  The gas around the
edge of the bubble has no velocity shift along the line of sight, 
whereas gas in the middle
of the bubble is being pushed both towards and away from the observer,
creating red- and blueshifted emission in the position-velocity diagram
in the middle of the cut through the bubble.  The spatial extent of the 
gas extension therefore defines the extent of the bubble.  In the continuum
image of Sgr A East the molecular gas extension lies just inside a larger 
 bubble in the eastern region of Sgr A East.  The bubble seen in the
continuum image
has a much larger physical extent than the bubble defined by the molecular
gas extension.  It is unclear whether these two features are related and
interacting with one another. \newline

\subsection{Southern SNR Interacting with the Molecular Ridge}

\ \ \ \ \ The morphology and kinematics of the northern and southern 
parts of the molecular ridge suggest different histories along
the length of the streamer.  Figure \ref{color.mom0} shows that
the morphology of the southern part of the streamer is 
fluffier and less confined than the northern part.  In the south
there are more protrusions and clumps in the gas, and the flux
level is lower than in the north.  While Figure \ref{22.11.mom0} 
indicates no clear temperature differences between the northern 
and southern parts, the velocity centroid image (Figure \ref{mom1})
 shows that the entire 
 southern half of the ridge is at a lower velocity than the northern half 
and that the transition between the two regions is not smooth.
The velocity dispersion map (Figure \ref{mom2}) reveals that the 
 northern half of the ridge has roughly the same dispersion throughout while 
the southern half has widely varying dispersions as well as 
regions of very high dispersion indicating perturbed gas.  
The position-velocity diagrams taken across the
ridge (Figure \ref{psvl2}) show a different gradient from east to
 west in the north and south.  The diagrams taken across the northern
part of the ridge show an east to west redshifting of the gas,  while
in the south the gradient blueshifts to the west.  \newline

\ \ \ \ \ The northern and southern parts of the ridge are spatially and 
kinematically connected but are clearly being acted upon by different forces.
The entire ridge is one structure, but currently there are two distinct
shells expanding into the northern and southern parts separately.
Sgr A East is expanding into the 50 \kms cloud, and the northern half
of the molecular ridge traces this impact.
There is apparently a SNR to the south of Sgr A East which is interacting
with the southern half of the ridge as well as Sgr A East and the southern
streamer. 
In terms of its galactic coordinates the SNR is ``G 359.92 -0.09''.  
Images of this SNR can be seen in Yusef-Zadeh and Morris (1987), 
in their Figures 2 and 3 of 20 cm continuum emission, 
which clearly show a circular feature south of Sgr A East.    
The SNR can also be seen in Pedlar et al. (1989), Figure 5b, another
20 cm continuum image of Sgr A East.  This SNR is interacting with 
Sgr A East, pushing up on it, creating the bend in the southern edge 
of Sgr A East seen in the color and grey-scale emission in 
Figures \ref{color.mom0} and \ref{bw.SNR.mom0}.  This region where
the SNR is apparently confining Sgr A East is exactly the location of the
1720 MHz OH maser group seen by Yusef-Zadeh et al. (1996) and 
Yusef-Zadeh et al. (1999).  Their
data show a group of 8 masers all positioned just inside the southern 
edge of Sgr A East, along the inward curve caused by the expanding SNR
to the south of Sgr A East.  Yusef-Zadeh et al. (1999) 
propose that these masers
are due to the expansion of Sgr A East into the 50 \kms cloud, as the
line of sight velocities of the masers are in the range 53 \kms to 66 \kms.
Our data suggest that the 50 \kms cloud lies mainly to the east of Sgr A East
and not to the south, as the molecular ridge does not continue along 
the southern boundary of Sgr A East.  However we find plenty of molecular
gas in the location of the masers, gas not associated with the 50 \kms cloud 
but rather with the southern streamer
as it transports gas from the 20 \kms cloud to the nuclear region.  The
masers appear to lie at the intersection of the SNR as it
expands into Sgr A East and the southern streamer as it moves towards 
the CND.  The masers may therefore arise from the expansion of the SNR
into the molecular gas in the southern streamer. \newline

\ \ \ \ \ The SNR is also 
interacting with the 20 \kms cloud, creating the sharp linear eastern
edge of the GMC seen in the contours of these same figures.  
The 20 \kms cloud does display a backwards ``C'' shape
structure in its position-velocity diagrams (Paper 1, Figure 8 cut b), 
indicating that this cloud is more behind the SNR than in front of it.  
The position-velocity diagrams of a cut taken along the lower 
edge of the cloud in Paper 1 (Figure 8 cut c ) show a forward ``C'' structure,
 with the gas in the middle of the cut being blueshifted along the line of
sight, where presumably the SNR is impacting the 20 \kms cloud and
disrupting the gas. This interaction may have created 
the southern streamer.  Our velocity
dispersion image (Figure \ref{mom2}) shows a pronounced increase
in line width in the 20 \kms gas located near the ``wisp''
in the continuum image.  The region with the increased line width has
an arc morphology, continuing along the bend of the ``wisp.''  This
area traces the impact of the expanding SNR on the 20 \kms cloud. \newline

\ \ \ \ \ This SNR appears to also be interacting with the southern half 
of the molecular ridge, causing the clear demarcation between the
northern and southern halves seen in the kinematics of the gas. 
The two halves of the ridge have different histories and 
expanding shells acting on them.  The interaction between the 
SNR and the southern part of the ridge can be seen in the 
position-velocity diagrams taken along the length of the ridge (see Figure
\ref{psvl1}).  These diagrams not only show large line widths for the 
emission in the south, but also display a hole in position-velocity space
 in the southern half of the diagrams between 35 \kms and 70 \kms.  This
may be a shell swept clear by the SNR expanding into the southern part
of the streamer.  This would place the bulk of the 
streamer in front of the SNR, and the heated high-velocity emission at 
$\geq$70 \kms would be redshifted emission from the streamer that is
being pushed away from the observer along the line of sight, on the 
far side of the SNR.  The U-shaped feature in the 1.3mm image
(Figure \ref{dust.mom0}) may well be a tilted circular swath of dust
 from the 20 \kms cloud wrapping around the SNR and projected into a 
U-shape along the line of sight, possibly connecting up with the southern
tip of the ridge.  The southern part of the ridge
may be material from the 20 \kms cloud that has been impacted by
the expanding SNR, pushing it away from the core of the GMC.
\newline

\ \ \ \ \ Given the masses of the ridge and the 20 \kms cloud, how feasible
is it that a SNR could move these features by the 20 \kms shift we observe
in the position-velocity diagrams?  Assuming equipartition of energy in the 
SNR, roughly half the 10$^{51}$ ergs released in the explosion would be
converted to kinetic energy.  Using the mass derived in Table 1 for the
southern part of the ridge, 0.5 10$^{51}$ ergs would be able to displace
gas in the southern ridge by 35 \kms.  Using the mass derived for the northern
part of the 20 \kms cloud as reported in Paper 1, a SNR
could shift the gas by 15 \kms.  These numbers are consistent with
the observed velocity shifts, so that it is entirely feasible for a typical
SNR to cause the observed velocity shifts and increased line widths.   
The northern part of the ridge shows a 20 \kms velocity shift due to the 
expansion of Sgr A East.  Using the derived mass of the northern part of
the ridge, we estimate that 4.5 10$^{50}$ ergs of kinetic energy would be 
required to cause this movement of the 50 \kms cloud, well within
the kinetic energy limit of a typical SNR.  \newline

\ \ \ \ \  It does not appear that the entire molecular ridge
was created by the expansion of Sgr A East into the 50 \kms cloud,
resulting in a narrow shocked region of gas. While the gas in
the northern part of the ridge traces the denser parts of the 50 \kms
cloud, including the western edge of the GMC that Sgr A East is expanding into,
the ridge continues
to the south past Sgr A East and is being acted upon by other
forces in this region.  The gas in the southern part of the ridge 
may have originated in the 20 \kms cloud and been pushed to its 
present position by the expanding SNR.  If the narrowness of the
ridge in the south is real, in the absence of apparent confinment mechanisms,
it could suggest a primordial filamentary structure.   \newline

\subsection{Line of Sight Locations of Features}

\ \ \ \ \ With our new data we are able to put constraints on
the line-of-sight locations of several of the features at the
Galactic Center and begin to build a 3-D model of the region.
Figure \ref{cartoon} is a schematic drawing of the large-scale
 features in the central 15 pc of the Galaxy, with positions
shown along the line of sight from the Sun where east is up and west
is down. 
The 20 \kms cloud is in front of the nuclear region as seen from the
redshift to the north in the velocity gradient along
the southern streamer (Paper 1).  This places the CND behind the 20
\kms cloud, and the 20 \kms cloud in front of the nuclear region.
Sgr A East is behind Sgr A West, as Sgr A West is seen in absorption
against Sgr A East at 90 cm (Figures 5 and 7 in Pedlar et al. 1989).
The 50 \kms cloud is to the east of and slightly behind Sgr A East,
as indicated in our position-velocity diagrams where there is redshifted
emission from the interaction with Sgr A East but no corresponding
blueshifted emission (Figure \ref{psvl1}).  
Along the molecular ridge there is a velocity 
gradient that redshifts to the north, placing the northern part of 
the ridge further away from the Sun along the line of sight than the southern
part.
The SNR to the south of Sgr A East is interacting with the southern part of
the molecular ridge which connects to the 50 \kms cloud to the north.  The
SNR is also interacting with the southern edge of Sgr A East, as well
as the eastern edge of the 20 \kms cloud.  The angular size of the SNR
is about 3.5$'$, which corresponds to roughly 8.5 pc (at a distance
of 8.5 kpc), so that Sgr A East and the 20 \kms cloud have to be within
8.5 pc of each other along the line of sight.      \newline

\subsection{Two Streamers as One Coherent Structure?}

\ \ \ \ \ The two streamers we image seem to be part of a larger, 
connected gas cloud complex,
as seen from the extended emission in the 1.3mm map and the connections
imaged by other groups between the two GMCs in the region.   We are 
missing zero-spacing information in our data, which essentially filters
out any structures greater than 1$'$ in size in our maps as well as causes
the negative trenches around the strongest emission.  In addition, 
NH$_3$ only traces the densest gas in the 
system, so we are not sensitive to less dense gas which may be connecting 
the streamers.  In order to understand the nature
of the large-scale structure of molecular gas at the Galactic Center, we 
must compare our maps to the single-dish 1.3mm image (assuming that the
dust and gas are well-mixed) and examine surveys of molecular clouds
in the region. G\"usten, Walmsley, \& Pauls (1981) 
argue that the 20 \kms cloud and 50 \kms
cloud are not only bound to each other but are condensations in a larger
complex of 5 gas clouds beginning in the south with the 20 \kms cloud
and extending almost 2 degrees north of the nucleus.  
The recent Nobeyama CS $J$=1-0 survey (Tsuboi, Handa \&
Ukita 1999) shows a large-scale ridge of gas extending along the Galactic
Plane for $\ge$0.5$^{\circ}$ at negative latitudes from Sgr A$^*$.  The
20 \kms cloud and 50 \kms cloud appear to be part of this large structure.
One obvious question about this structure is why it is offset to negative 
latitudes and why there is little corresponding emission at positive latitudes.
\newline

\ \ \ \ \ It is possible that the two streamers we image together make
up a tilted ring with respect to our line of sight.
A ring-like structure if it were in the plane of the Galaxy would
appear as a linear structure from our point of view in the disk, so that 
morphology alone would not
necessarily indicate the presence of a ring; kinematic evidence is needed.
Our velocity coverage is limited, sensitive to emission with radial 
velocities of -35 \kms to 100 \kms.  The Nobeyama survey paper (\cite{tsu99})
presents a position-velocity diagram (their Figure 2b) centered on Sgr A$^*$
summed over 16$'$ in latitude which shows half of a ring-type structure
in the central 0.5$^{\circ}$ with emission at positive velocities ranging
from 0 \kms to 90 \kms.  There is a small region of emission at -20 \kms
opposing this semi-complete ring which may complete the circle.  \newline

\section{Conclusions}

\ \ \ \ \ A VLA 5-field mosaic of the inner 15 pc of the 
Galaxy in \11 and \22 emission reveal two long, narrow streamers 
running roughly along the Galactic plane, each
tracing connections between continuum emission in the nucleus and
nearby GMCs.  The southern streamer is carrying gas from the 20 \kms cloud 
to the CND (see Paper 1).
The other streamer, called the molecular ridge, traces the dense parts of 
the 50 \kms cloud, including the core and much of the southern part of
the GMC, and wraps around the eastern edge of Sgr A East.  The expansion of 
Sgr A East into the 50 \kms cloud is seen as a ``C''-shaped structure in 
position-velocity diagrams of the northern part of the molecular ridge. 
There is a velocity gradient along the length of the entire streamer with
the gas being redshifted to the north, similar to the gradient
seen in the southern streamer connecting the 20 \kms cloud to the CND.  
The molecular ridge closely follows the arc of the Sgr A East 
continuum emission and continues south past Sgr A East towards the
20 \kms cloud.  The gas in the southern part of the ridge is perturbed, 
showing heating and increased line widths, likely the result of an 
interaction with a SNR south of Sgr A East.  This SNR appears to be 
impacting the 20 \kms cloud and Sgr A East as well, 
placing these features within
10 pc of each other along the line of sight.  Our velocity information
allows us to place the 20 \kms cloud in front of the CND and Sgr A West,
as well as the 50 \kms cloud behind Sgr A East.  These two streamers
appear to be dense peaks of a larger gas complex surrounding the nuclear
region.  \newline

\ \ \ \ \ We would like to thank Cornelia Lang for lending us
her VLA 20 cm continuum image and Robert Zylka for sharing
 his IRAM 1.3mm image.

\newpage

\figcaption[final_image/1.cartoon.ps]{A schematic drawing of the relevant
 large-scale features in the central 15 pc of the Galaxy showing positions
on the plane of the sky with east to the left. \label{cartoon1}}

\figcaption[final_images/color.mom0.sgra.cps]{Velocity-integrated maps 
of \11 and
\22 emission in contours from the central 10 by 15 pc of the Galaxy 
overlaid on a color 20 cm continuum image of Sgr A East and West.  
Two north-south streamers are seen, one connecting the 20 \kms cloud
in the south to the nuclear region, and the other tracing the eastern
edge of Sgr A East where it interacts with the 50 \kms cloud  and 
continuing south towards the 20 \kms cloud.
The contours levels are -2, -1, 0.1, 1, 2, 3, 4, 5, 6, 7, 8, 9, 10,
 11,  and 12 Jy beam$^{-1}$ \kms, integrated over 135 \kms.
Five 2$'$ fields have been mosaiced together, and only emission corresponding
to a beam response of $\geq$15\% has been plotted here.  The 14.5$''$ by 
8.8$''$ beam is shown in the lower left corner of the \11 map. The color
continuum image ranges from 0 to .18 Jy beam$^{-1}$. \label{color.mom0}}

\figcaption[final_images/beam.pattern.mom0.ps]{The 5-field 
mosaiced primary beam is seen in dotted contours, with 10\% response levels 
labeled. The \11 velocity-integrated map from Figure 1 is seen
in grey-scale and solid contours.  \11 emission below a 15\% primary beam
level has not been shown.   \label{beam.mom0}}

\figcaption[final_images/bw.mom0.SNR.ps]{Same as Figure \ref{color.mom0} 
with grey-scale ranging from 0 to .028 Jy beam$^{-1}$. \label{bw.SNR.mom0}}

\figcaption[final_images/sgra12.mom0.ps]{Velocity-integrated maps
of \11 emission from the southeastern field only on the left and the
southwestern field only on the right.  No mosaicing was used in these
images, so that only data from one pointing is shown in each map.  The
contours are the same as in Figure 1.  \label{sgra12.mom0}}

\figcaption[final_images/sidelobe.fig.ps]{Position-velocity diagrams 
investigating possible sidelobe contamination in the southern part of
the ridge due to the 20 \kms cloud.  Cut A shows the
kinematics along the ridge, while cut B corresponds to the 20 \kms
cloud. The contours are 0.04, 0.08, 0.12, 0.16, 0.20, 0.24, 0.28, 0.32, 
0.36 and 0.40 Jy beam$^{-1}$ for cut A and 0.1, 0.2, 0.3, 0.4, 0.5, 0.6, 0.7,
0.8, 0.9 and 1.0 Jy beam$^{-1}$ for cut B.   \label{sidelobe}}

\figcaption[final_images/dust.mom0.ps]{A 1.3mm continuum image of the
central 17.5 by 25 pc of the Galaxy  is shown in contours (Zylka 1997),
 tracing free-free and thermal dust emission.  The contour levels are
0.32, 0.47, 0.65, 0.85, 1.20, 1.65, 2.50, and 3.50 Jy beam$^{-1}$ and the beam
size is $\sim$11$''$.  Our \11 velocity-integrated map is
shown in grey-scale from 0 to 8 Jy beam$^{-1}$ \kms. \label{dust.mom0}}

\figcaption[final_images/22.11.mom0.ps]{\22/\11 velocity-integrated emission
is shown in grey-scale with the ratio ranging from 0 to 2. The ratio of
\22 to \11 flux reflects heating of the gas, where the dark regions trace
heated gas.
20 cm continuum emission from Sgr A East and West is overlaid in contours
corresponding to 0.04, 0.06, 0.08, 0.10, 0.12, 0.14, 0.16, 0.18, 
0.20 Jy beam$^{-1}$.  \label{22.11.mom0}}

\figcaption[final_images/mom1.fig.ps]{Velocity centroid maps for both the
\11 and \22 emission are shown in grey-scale with continuum emission from
Sgr A East and West in contours.  The grey-scale ranges from 15 \kms in
white to 60 \kms in black, and the 20 cm continuum emission contours are 
the same as in 
Figure \ref{22.11.mom0}. A velocity gradient is evident along the
length of the eastern streamer, where the gas is redshifted in the north
where it wraps around Sgr A East.  \label{mom1}}

\figcaption[final_images/mom2.fig.ps]{Velocity dispersion maps for
the \11 and \22 data in grey-scale ranging from 0 \kms to 25 \kms. The
20 cm continuum emission contours are the same as in Figure \ref{22.11.mom0}.
 \label{mom2}}

\figcaption[final_images/psvl.long.ps]{Position velocity diagrams for
cuts running north to south along the molecular ridge are shown for the \11
and \22 datacubes, as well as the divided \22/\11 emission which traces
heated gas.  The contour levels are 0.04, 0.08, 0.12, 0.16, 0.20, 0.24, 0.28, 
0.32, 0.36, 0.40, 0.44, 0.48, 0.52, 0.56 and 0.60 Jy beam$^{-1}$ for the \11
and \22 diagrams.  The \22/\11 diagrams have a grey-scale ranging from 0 
to 2 and the contour levels are 0.5, 1.0, 1.5, and 2.0.   \label{psvl1}}

\figcaption[final_images/psvl.top.ps]{Position velocity diagrams
for cuts running east to west across the molecular ridge are shown
for the \11, \22 and \22/\11 datacubes.  The contour levels are 
0.0250, 0.0625, 0.1000, 0.1375, 0.1750, 0.2125, 0.2500 , 0.2875, 0.3250, 
0.3625, 0.4000, 0.4375, 0.4750, 
 0.5125 and 0.5500  Jy beam$^{-1}$ for the \11 and \22 diagrams.  The \22/\11 
diagrams have a grey-scale ranging from 0 to 2 and the contour levels 
are 0.5, 1.0, 1.5, and 2.0.   \label{psvl2}}

\figcaption[final_images/GCcartoon.ps]{A schematic drawing of the 
large-scale features in the central 15 pc of the Galaxy
 showing positions along the line of sight from the Sun with east being up. 
\label{cartoon}}

\begin{deluxetable}{lcccccccc}
\tablecolumns{9}
\tablecaption{NH$_3$(1,1) Derived Physical Parameters for the Molecular Ridge \label{tab1}} 
\tablehead{ 
\colhead{Peak} & \colhead{R.A.; Dec.} & \colhead{$\tau_m$} & \colhead{$\Delta T_a^*$}  & \colhead{$T_{ex}$}  & \colhead{$T_{rot}$}  &  \colhead{$N(J,K)$} & \colhead{$N$(H$_2$)} & \colhead{Mass}
\\ \colhead{ } & \colhead{(B1950)} & \colhead{} & \colhead{$(K)$} & \colhead{$(K)$} & \colhead{$(K)$} & \colhead{($10^{15} cm^{-2}$)}  & \colhead{($10^{24} cm^{-2}$)} & \colhead{($10^5 M_{\odot}$)}
}
\startdata
Northern & 17:42:40; -28:59:13 & 2.0 & 4.3 & 7.7 & 20 &
5.5 & 1.3 & 1.1 \\  
Southern & 17:42:37; -29:01:23 & 1.5 & 2.8 & 6.3 & 30 &
2.2 & 0.52 & 0.40 \\
\enddata
\end{deluxetable}

\end{document}